\documentclass[
    acmtog, %
    authorversion=true,
    screen=true,
    authordraft=false,
    review=false,
    timestamp=false,
    anonymous=false, 
    nonacm=true]{acmart}

\usepackage{tabularx}

\AtBeginDocument{%
	\providecommand\BibTeX{{%
			\normalfont B\kern-0.5em{\scshape i\kern-0.25em b}\kern-0.8em\TeX}}}

\author{Haotian Yin}
\affiliation{
    \institution{New Jersey Institute of Technology}      
    \country{United States}             
}
\email{hy9@njit.edu}

\author{Aleksander Plocharski}
\affiliation{
    \institution{Warsaw University of Technology}    
    \and 
    \institution{IDEAS NCBR}
    \country{Poland}
}
\email{aleksander.plocharski@pw.edu.pl}
            
\author{Michal Jan Wlodarczyk}
\affiliation{
    \institution{Warsaw University of Technology}
    \country{Poland}
}
\email{mwlodarzc@gmail.com}

\author{Przemyslaw Musialski}
\affiliation{
    \institution{New Jersey Institute of Technology}           
    \country{United States}             
}
\email{przem@njit.edu}

\acmSubmissionID{185}

\begin{document}

\title{A Finite Difference Approximation of Second Order Regularization of Neural-SDFs}

\begin{abstract}
We introduce a finite-difference framework for curvature regularization in neural signed distance field (SDF) learning. Existing approaches enforce curvature priors using full Hessian information obtained via second-order automatic differentiation, which is accurate but computationally expensive. Others reduced this overhead by avoiding explicit Hessian assembly, but still required higher-order differentiation. In contrast, our method replaces these operations with lightweight finite-difference stencils that approximate second derivatives using the well known Taylor expansion with a truncation error of \(O(h^2)\), and can serve as drop-in replacements for Gaussian curvature and rank-deficiency losses. Experiments demonstrate that our finite-difference variants achieve reconstruction fidelity comparable to their automatic-differentiation counterparts, while reducing GPU memory usage and training time by up to a factor of two. Additional tests on sparse, incomplete, and non-CAD data confirm that the proposed formulation is robust and general, offering an efficient and scalable alternative for curvature-aware SDF learning.
\end{abstract}

\begin{teaserfigure}
    \centering
    \includegraphics[width=0.99\linewidth,trim=0 42 0 0,clip]{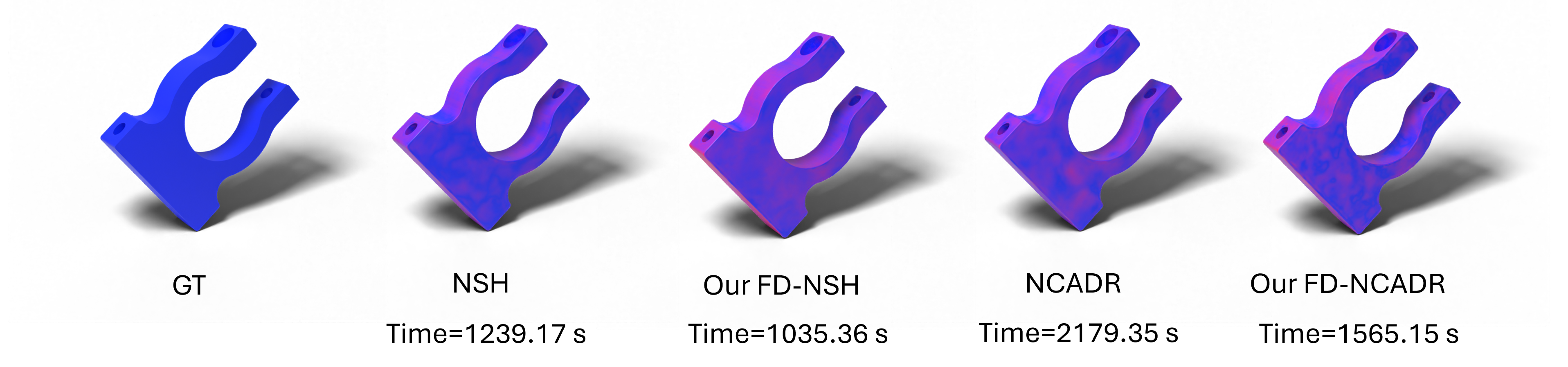}
\caption{Neural-SDF reconstructions with Hausdorff–distance error maps (blue $\rightarrow$ low, red $\rightarrow$ high) of baseline methods NSH~\shortcite{Wang2023NeuralSingularHessianIN} and NeurCADRecon~\shortcite{Dong2024NeurCADReconNR} compared to our proposed FD-counterparts. 
}
    \label{fig:hausdorff}
\end{teaserfigure}

\maketitle

\section{Introduction}
Neural signed distance fields (SDFs) are a widely used representation for surface learning and reconstruction. For CAD-like objects, however, standard first-order regularizers such as the Eikonal loss are insufficient to capture sharp features and developable regions. Methods like NeurCADRecon (NCR)~\cite{Dong2024NeurCADReconNR} and Neural-Singular-Hessian (NSH)~\cite{Wang2023NeuralSingularHessianIN} addressed this limitation by introducing curvature-based losses derived from second-order automatic differentiation. While effective, these approaches are computationally demanding, as they require explicit evaluation of the Hessian or its determinant during training.

FlatCAD~\cite{Yin2025FlatCAD}  proposed a more efficient alternative by formulating a curvature regularizer that avoids assembling the full Hessian. This reduces overhead compared to full Hessian methods, yet still incurs the cost of higher-order automatic differentiation. 

FlatCAD also sketched an initial variant based on finite differences. 
We build on this idea and introduce a general finite-difference (FD) formulation for curvature regularization in neural SDFs. By combining a few forward evaluations with first-order gradients, our method enforces curvature constraints without requiring any second-order automatic differentiation. 
Recently, a rendering approach applied a similar forward-evaluation principle for higher-order estimation~\cite{wang2024stochastic}, underpinning that concept.

Our scheme is second-order accurate, framework-agnostic, and computationally significantly cheaper than auto-diff methods. 
Our contributions are: 
  (1) A finite-difference discretizations of curvature regularizers for neural SDFs that achieve similar accuracy using only first-order terms. 
  (2) An implementation that avoids higher-order auto-diff entirely, requiring only forward SDF evaluations. 
  (3) Empirical validation on the ABC dataset showing that our method matches or improves the reconstruction quality SOTA, while reducing memory usage and training time by up to \(2\times\).

\begin{figure*}[t]
    \centering
    \includegraphics[width=0.99\linewidth,trim=0 10 0 50,clip]{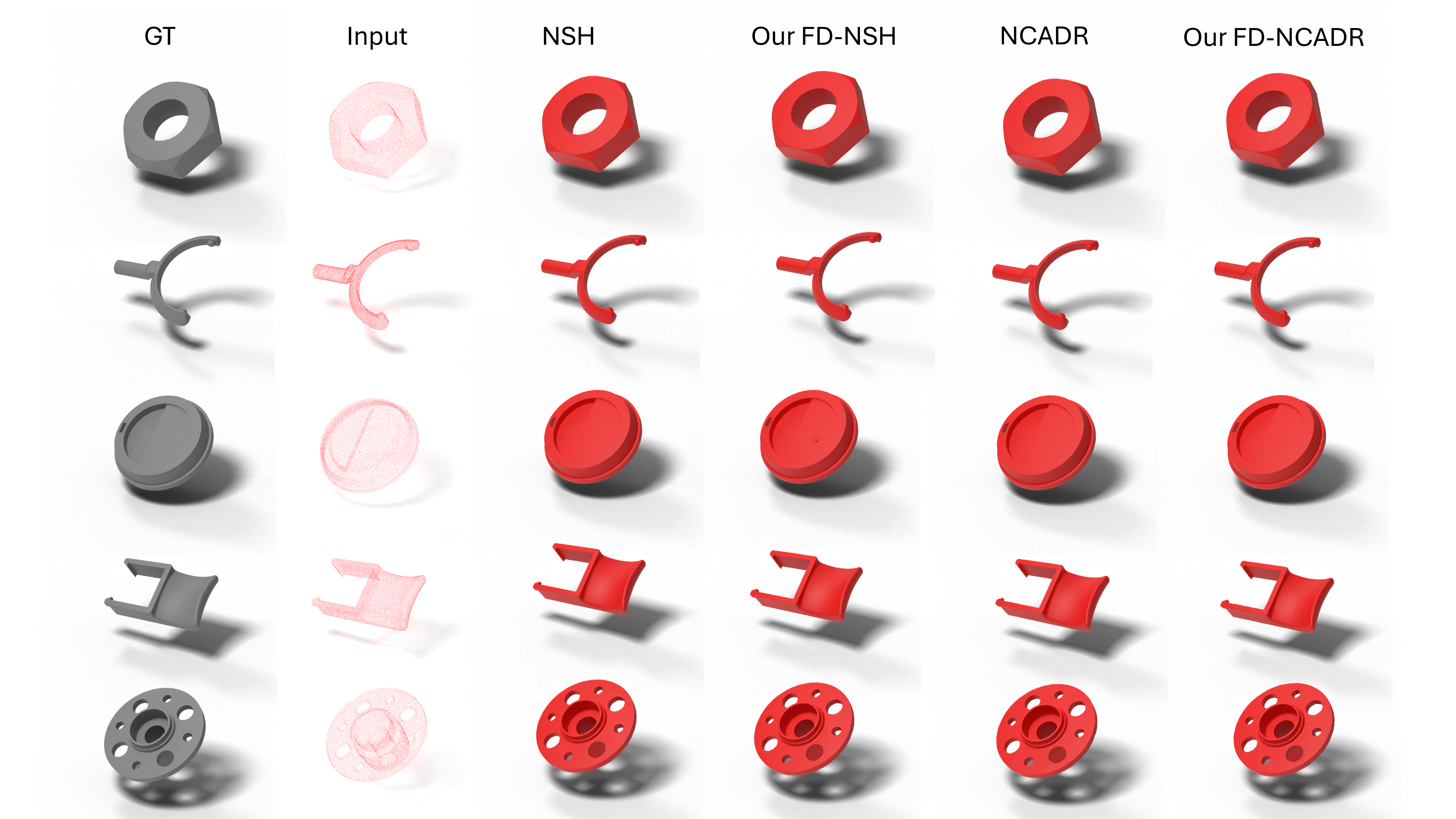}
    \caption{Comparison with state-of-the-art surface reconstruction. FD and AD proxies yield comparable accuracy. Our FD approach avoids second-order differentiation while maintaining efficiency.}
    \label{fig:ablation_finite}
\end{figure*}

\section{Related Work}\label{sec:related}

Implicit functions are a widely used representation for 3D reconstruction. Classical methods estimate signed distances from tangent planes~\cite{Hoppe1992SurfaceRF}, radial basis functions~\cite{Carr2001ReconstructionAR}, or Poisson surface reconstruction~\cite{Kazhdan2006PoissonSR}. Neural implicit methods define surfaces as the zero level-set of a learned field, with DeepSDF~\cite{Park2019DeepSDFLC} as a seminal example. Self-supervised approaches such as IGR~\cite{Gropp2020ImplicitGR} and SAL~\cite{Atzmon2019SALSA} introduced Eikonal and sign-agnostic losses, while SIREN~\cite{Sitzmann2020ImplicitNR} improved high-frequency fitting.

Beyond first-order constraints, recent work has emphasized curvature-aware priors. DiGS~\cite{BenShabat2021DiGSD} penalizes divergence to promote smoothness, while other methods introduced curvature or rank-based regularizers for developability~\cite{Novello_2023_ICCV,Selvaraju2024_Developa}. Neural-Singular-Hessian (NSH)~\cite{Wang2023NeuralSingularHessianIN} minimizes the smallest singular value of the Hessian to enforce rank-deficiency, and NeurCADRecon~\cite{Dong2024NeurCADReconNR} uses a Gaussian curvature loss derived from the Hessian. FlatCAD~\cite{Yin2025FlatCAD,Yin2025_Scheduling} proposed a more efficient determinant-based regularizer that avoids explicit Hessian assembly but still relies on higher-order derivatives. These methods improve fidelity and feature preservation but suffer from the computational burden of second-order derivatives.

\section{Our Approach}

We propose a finite-difference regularization framework for learning neural signed distance fields (SDFs) from unoriented point clouds. Following the standard self-supervised setup for neural implicit reconstruction, we combine a point-based Dirichlet condition and an Eikonal loss to ensure surface fidelity and valid SDF behavior. Our core contribution is a second-order regularizer based on finite differences. Instead of computing Hessians via costly second-order automatic differentiation, we approximate directional second derivatives from local SDF evaluations using Taylor expansion stencils. This enables lightweight reformulations of curvature-based penalties such as those in NRC~\cite{Dong2024NeurCADReconNR} and NSH~\cite{Wang2023NeuralSingularHessianIN}, while achieving $O(h^2)$ accuracy and avoiding full second-order computation graphs.

\subsection{Differential-Geometric Background}

A surface $\mathcal{M}\subset\mathbb{R}^{3}$ is represented as the zero-level set of a signed distance field (SDF): 
\[
f: \mathbb{R}^{3}\rightarrow \mathbb{R},\quad \mathcal{M}=\{\mathbf{x}\mid f(\mathbf{x})=0\}.
\]
A valid SDF satisfies the Eikonal condition $\|\nabla f\|=1$ in a neighborhood around $\mathcal{M}$, implying zero second derivative along the normal:
\[
\partial_{nn}f=\mathbf{n}^{\top}H_f\,\mathbf{n}=0,\quad \mathbf{n}=\tfrac{\nabla f}{\|\nabla f\|},
\]
where $H_f$ denotes the Hessian. Projecting $H_f$ onto an orthonormal tangent basis $(\mathbf{u},\mathbf{v})$ yields the shape operator
\[
S_\mathcal{M}=
\begin{pmatrix}
\mathbf{u}^{\!\top}H_f\,\mathbf{u} & \mathbf{u}^{\!\top}H_f\,\mathbf{v}\\
\mathbf{v}^{\!\top}H_f\,\mathbf{u} & \mathbf{v}^{\!\top}H_f\,\mathbf{v}
\end{pmatrix},
\]
whose eigenvalues are the principal curvatures $\kappa_1,\kappa_2$ with Gaussian curvature $K=\kappa_1\kappa_2=\det(S_\mathcal{M})$. Any uniformly random orthogonal completion $\mathbf{u}$, $\mathbf{v}$ of the gradient normal is equally valid, and expectation over such random frames recovers the correct curvature measure~\cite{Yin2025FlatCAD}. 

To avoid second-order automatic differentiation, we approximate the entries of $S$ with finite-difference stencils. For step size $h>0$ and tangent directions $\mathbf{u},\mathbf{v}$, the second derivatives are approximated as
\[
f_{uu} \approx \frac{f(\mathbf{x}_0+h\mathbf{u}) - 2f(\mathbf{x}_0) + f(\mathbf{x}_0-h\mathbf{u})}{h^2},
\]
\[
f_{vv} \approx \frac{f(\mathbf{x}_0+h\mathbf{v}) - 2f(\mathbf{x}_0) + f(\mathbf{x}_0-h\mathbf{v})}{h^2},
\]
\[
f_{uv} \approx \tfrac{1}{4h^2}\Big(f(\mathbf{x}_0{+}h\mathbf{u}{+}h\mathbf{v}) - f(\mathbf{x}_0{+}h\mathbf{u}{-}h\mathbf{v})
 - f(\mathbf{x}_0{-}h\mathbf{u}{+}h\mathbf{v}) + f(\mathbf{x}_0{-}h\mathbf{u}{-}h\mathbf{v})\Big).
\]
By Taylor expansion, these approximations incur an error of $O(h^2)$.

\begin{figure}[b]
    \centering    \includegraphics[width=0.99\linewidth,trim=0 90 0 130,clip]{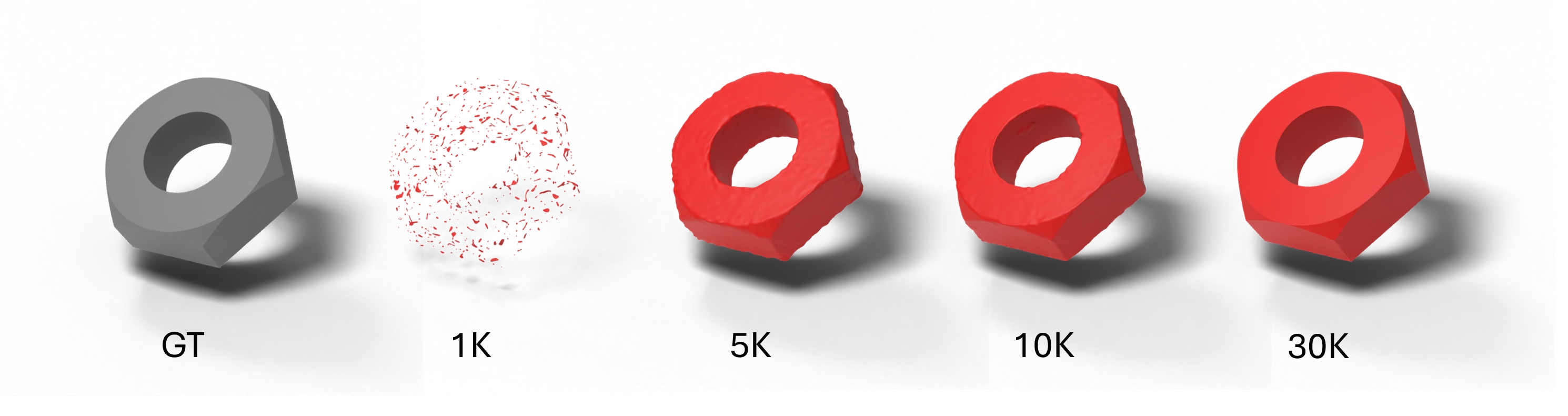}
    \caption{Qualitative reconstructed results on point clouds with varying levels of sparsity.}
    \label{fig:sparcity}
\end{figure}

\subsection{FD-NeurCADRecon}
NeurCADRecon (NRC)~\cite{Dong2024NeurCADReconNR} enforces piecewise-developability by penalizing the Gaussian curvature of the learned SDF. In the original formulation, this is expressed via the Hessian $H_f$ as
\[
K = - \frac{\det\begin{pmatrix} 
H_f(\mathbf{x}_0) & \nabla f(\mathbf{x}_0)^\top \\
\nabla f(\mathbf{x}_0) & 0 
\end{pmatrix}}{\|\nabla f(\mathbf{x}_0)\|^4}.
\]
Evaluating this term requires explicit Hessian–vector products through second-order automatic differentiation, which is costly in both runtime and memory.

In our finite-difference (FD) formulation, we avoid second-order derivatives entirely. We first construct a local tangent frame $(\mathbf{u},\mathbf{v})$ at an off-surface shell point $\mathbf{x}_0$, and approximate the directional second derivatives $f_{uu}, f_{vv}, f_{uv}$ via central-difference stencils of nearby function evaluations. Gaussian is then estimated as
\[
K_{\text{FD}}(\mathbf{x}_0) \;=\; \frac{f_{uu} f_{vv} - f_{uv}^2}{\|\nabla f(\mathbf{x}_0)\|^4},
\]
which is second-order accurate by Taylor expansion with truncation error $O(h^2)$. This replacement yields the same geometric bias as the original NRC loss while reducing both memory consumption and wall-clock training time.

\subsection{FD-Neural-Singular-Hessian}
The Neural-Singular-Hessian (NSH) regularizer~\cite{Wang2023NeuralSingularHessianIN} promotes developability by minimizing $\det(H_f)$ near the surface, thereby forcing the Hessian to be rank-deficient. This suppresses spurious curvature and aligns normals in a piecewise-developable manner. The drawback is that explicit determinant evaluation requires full Hessian computation.

We reformulate this regularizer in the tangent frame $(\mathbf{u},\mathbf{v})$ using finite differences. Specifically, the projected determinant is approximated as
\[
D_{\text{FD}}(\mathbf{x}_0) \;=\; f_{uu} f_{vv} - f_{uv}^2,
\]
where $f_{uu}, f_{vv}, f_{uv}$ are obtained via the same central-difference stencils as above. By construction,
\[
D_{\text{FD}}(\mathbf{x}_0) = (\mathbf{u}^\top H_f \mathbf{u})(\mathbf{v}^\top H_f \mathbf{v}) - (\mathbf{u}^\top H_f \mathbf{v})^2 + O(h^2),
\]
guaranteeing second-order consistency. As $D_{\text{FD}}$ approaches zero, at least one principal curvature vanishes, ensuring zero normal curvature while preserving sharp features. This finite-difference formulation thus retains the geometric effect of NSH but removes the overhead of second-order automatic differentiation.

\subsection{Practical Loss Implementation}
In practice, we leverage the property $\|\nabla f\|\approx 1$ near the surface, simplifying denominators in curvature terms. We found that early training gradients introduce oscillations, and in FD-NRC we replace $\|\nabla f(\mathbf{x}_0)\|$ by $1$ for near-surface points for stability. %

Our total loss is
\begin{equation}
\mathcal{L}_{\text{total}}
  = \mathcal{L}_{\text{DM}}
  + \lambda_{\text{DNM}}\mathcal{L}_{\text{DNM}}
  + \lambda_{\text{eik}}\mathcal{L}_{\text{eik}}
  + \lambda_{\text{fd}}\mathcal{L}_{\text{fd}},
\end{equation}
where $\mathcal{L}_{\text{DM}}$ is a Dirichlet condition anchoring input points to the zero-level set, $\mathcal{L}_{\text{DNM}}$ a non-manifold penalty following~\cite{Atzmon2020SALSA}, $\mathcal{L}_{\text{eik}}$ the Eikonal constraint~\cite{Gropp2020ImplicitGR}, and $\mathcal{L}_{\text{fd}}$ our finite-difference curvature regularizer.

\section{Experiments and Results} \label{sec:experiments}

We validate our finite-difference regularizers FD-NCR and FD-NSH against their automatic-differentiation counterparts NeurCADRecon (NRC)~\shortcite{Dong2024NeurCADReconNR} and Neural-Singular Hessian (NSH)~\shortcite{Wang2023NeuralSingularHessianIN}.

\paragraph{Datasets.}  
Experiments are performed on two subsets of the ABC dataset~\cite{Koch_2019_CVPR} to assess accuracy, robustness, and efficiency.
These subsets of 100 shapes each are: a pseudo-random \emph{1 MB set}, and a curated \emph{5 MB set} (3.5–9.5 MB average) with clean topology and sharp features. For each mesh, we sample 30k surface points; 20k are used per iteration, with 20k off-surface samples drawn uniformly in the bounding box.

\paragraph{Methods and Setup.}  
All methods are trained in a unified framework with a 4-layer SIREN MLP~\cite{Sitzmann2020ImplicitNR} (256 hidden units), Adam optimizer ($5\times10^{-5}$), and early stopping if Chamfer Distance does not improve for 1500 iterations. Hyperparameters follow the originals. 
 
Runtime results on the 1 MB set are measured on an NVIDIA H100 GPU. Qualitative and ablation studies additionally use A100, L4, and T4 GPUs; only H100 numbers are reported for consistency.

We report Chamfer Distance (CD, $\times10^{3}$, lower is better), F1 Score (F1, $\times10^{2}$, higher is better), and Normal Consistency (NC, $\times10^{2}$, higher is better), averaged over 100 shapes per subset. On the 1 MB set we also measure GPU memory, per-iteration time, and convergence time (best iteration $\times$ iteration time).

\paragraph{Quantitative and Qualitative.}  
FD-NCR and FD-NSH achieve accuracy on par with their AD versions, while reducing memory and runtime (Table~\ref{tab:acc_eff} and supplement). Hausdorff heatmaps (Fig.~\ref{fig:hausdorff}) show reduced local errors, and reconstructions (Fig.~\ref{fig:ablation_finite}) confirm comparable fidelity to NRC and NSH.

\begin{figure}[t]
    \centering \includegraphics[width=0.99\linewidth,trim=0 95 0 130,clip]{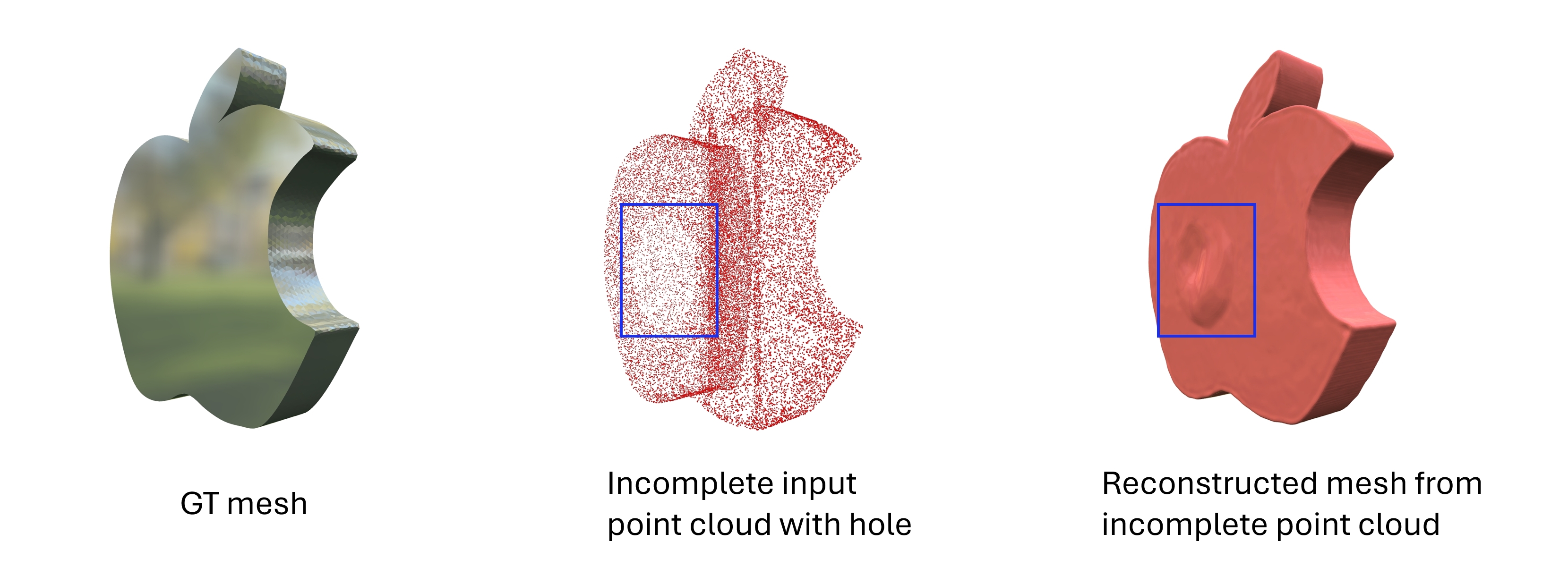}
    \caption{Reconstruction from incomplete input clouds. Missing planar regions are plausibly filled while preserving global continuity.}
    \label{fig:incomplete}
\end{figure}

With reduced sampling (30k $\to$ 10k, 5k, 1k points), our method maintains quality at 10k and 5k; only 1k shows strong degradation (Fig.~\ref{fig:sparcity}, Table: supplement). With holes excised from the input, geometry degrades gracefully: CD increases by $\sim$64\% and NC drops by $0.7\%$, while global topology is preserved(Fig.~\ref{fig:incomplete}, Table: supplement).
On non-CAD data (e.g., Armadillo), FD-NCR reconstructs smooth, topologically correct surfaces comparable to NRC but at about half the runtime (cf. supplement for figures and table).

\vspace{-5pt}
\section{Conclusion}
We presented a finite-difference framework for curvature regularization of neural signed distance fields. By approximating second derivatives with FD-stencils, our method avoids higher-order automatic differentiation while retaining second-order accuracy. 
We demonstrated results on two SOTA methods that achieve comparable or improved reconstruction quality with substantially lower memory and runtime cost. Future work includes extending the framework to adaptive sampling strategies and exploring applications beyond surface reconstruction.
\noindent
Please refer to supplemental material for more quantitative results.

\begin{figure}[b]
    \centering
    \includegraphics[width=0.99\linewidth,trim=0 140 0 120,clip]{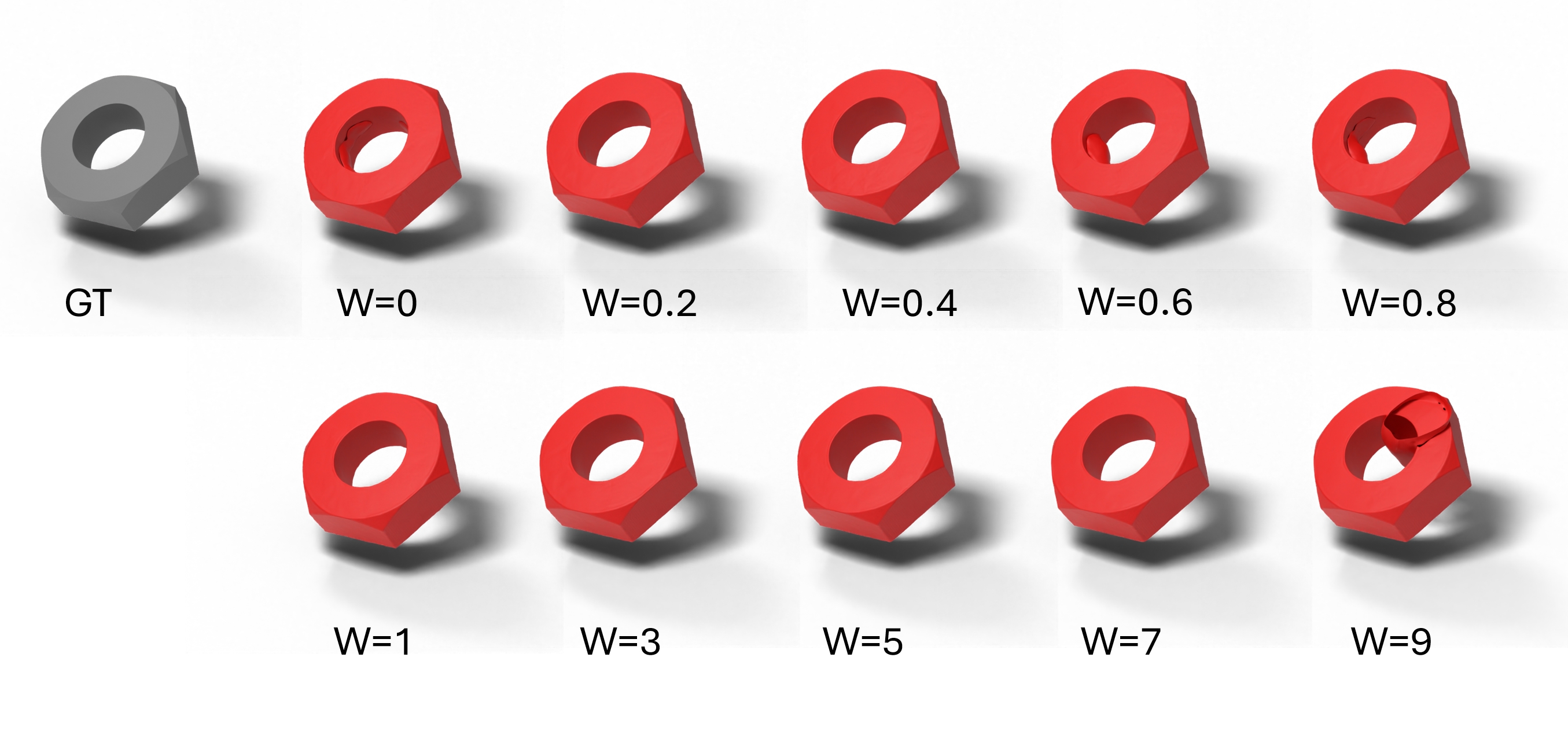}
    \caption{Reconstructions with different regularization weights. Results remain stable across a wide range.}
    \label{fig:ablation_weights}
\end{figure}

\begin{table}[b]
\centering
\small
\caption{Accuracy and efficiency (full table in supplamental).}
\label{tab:acc_eff}
\vspace{1mm}
\renewcommand{\arraystretch}{1.1}
\setlength{\tabcolsep}{3pt}
\begin{tabular*}{\columnwidth}{@{\extracolsep{\fill}}lcccccc@{}}
\toprule
Method & NC~$\uparrow$ & CD~$\downarrow$ & F1~$\uparrow$ & Conv.\ [s] & Mem.\ [GB] \\
\midrule
NSH~\cite{Wang2023NeuralSingularHessianIN} 
    & {93.93} & \textbf{{2.74}} & \textbf{91.91} & 559.41 & 6.09 \\
NSH-FD (Our)
    & \textbf{94.96} & 2.93 & 89.90      &  \textbf{362.95} & \textbf{4.30} \\
\midrule
NCR~\cite{Dong2024NeurCADReconNR} 
    & \textbf{93.71} & \textbf{2.65} & \textbf{90.63} &  391.23 & 6.06 \\
NCR-FD (Our)
    & 93.41 & 4.10 & 84.78 & \textbf{331.29} & \textbf{4.03} \\
\bottomrule
\end{tabular*}
\end{table}

\appendix

\section{Taylor Series Discretization}  

The derivation of discrete forms from continuous-time dynamics using Taylor series expansions~\cite{taylor1715methodus} is a foundational technique in numerical analysis and computational mathematics. This approach enables the transformation of differential equations into discrete-time approximations suitable for digital implementation. The core methodology relies on the Taylor series representation of a function \( f(x) \), where higher-order derivatives provide local approximations:  
\[
f(x_0 + h) \approx f(x_0) + h f'(x_0) + \frac{h^2}{2} f''(x_0) + \cdots + \frac{h^n}{n!} f^{(n)}(x_0)
\]  
Taylor series–based discretization is a well-established technique in numerical analysis and dynamical systems, where it is used to approximate differential quantities at specific locations with controlled accuracy. In the context of implicit surface learning, this approach has inspired finite-difference approximations for second-order geometric terms. By expanding the neural SDF around a point and evaluating it at nearby offsets, one can construct lightweight finite stencils to estimate Hessian entries and curvature measures without computing full second-order derivatives via automatic differentiation. In our work, we apply Taylor expansion techniques to discretize the regularization terms originally used in NSH~\cite{Wang2023NeuralSingularHessianIN} and NeurCADRecon~\cite{Dong2024NeurCADReconNR}, providing finite-difference formulations for the Hessian and curvature-based losses. This significantly improves computational efficiency by avoiding costly second-order automatic differentiation.

\subsection{Differential-Geometric Background}
A surface $\mathcal{M}\subset\mathbb{R}^{3}$ is represented as the zero-level set of a signed distance field (SDF):
\[
f: \mathbb{R}^{3}\rightarrow \mathbb{R},\quad \mathcal{M}=\{\mathbf{x}\mid f(\mathbf{x})=0\}.
\]
A valid SDF satisfies the Eikonal condition $\|\nabla f\|=1$ in a neighborhood around $\mathcal{M}$, implying zero second derivative along the normal:
\[
\partial_{nn}f=\mathbf{n}^{\top}H_f\,\mathbf{n}=0,\quad \mathbf{n}=\tfrac{\nabla f}{\|\nabla f\|},
\]
where $H_f$ denotes the Hessian. Projecting $H_f$ onto an orthonormal tangent basis $(\mathbf{u},\mathbf{v})$ yields the shape operator
\[
S_\mathcal{M}=
\begin{pmatrix}
\mathbf{u}^{\!\top}H_f\,\mathbf{u} & \mathbf{u}^{\!\top}H_f\,\mathbf{v}\\
\mathbf{v}^{\!\top}H_f\,\mathbf{u} & \mathbf{v}^{\!\top}H_f\,\mathbf{v}
\end{pmatrix},
\]
whose eigenvalues are the principal curvatures $\kappa_1,\kappa_2$ with Gaussian curvature $K=\kappa_1\kappa_2=\det(S_\mathcal{M})$. Any uniformly random orthogonal completion $\mathbf{u}$, $\mathbf{v}$ of the gradient normal is equally valid, and expectation over such random frames recovers the correct curvature measure~\cite{Yin2025FlatCAD}. 

To avoid second-order automatic differentiation, we approximate the entries of $S$ with finite-difference stencils. For step size $h>0$ and tangent directions $\mathbf{u},\mathbf{v}$, the second derivatives are approximated as
\[
f_{uu} \approx \frac{f(\mathbf{x}_0+h\mathbf{u}) - 2f(\mathbf{x}_0) + f(\mathbf{x}_0-h\mathbf{u})}{h^2},
\]
\[
f_{vv} \approx \frac{f(\mathbf{x}_0+h\mathbf{v}) - 2f(\mathbf{x}_0) + f(\mathbf{x}_0-h\mathbf{v})}{h^2},
\]
\[
\begin{aligned}
f_{uv}(\mathbf{x}_0) \approx \frac{1}{4h^2}\Big(
&\, f(\mathbf{x}_0 + h\mathbf{u} + h\mathbf{v}) - f(\mathbf{x}_0 + h\mathbf{u} - h\mathbf{v}) \\
&\, - f(\mathbf{x}_0 - h\mathbf{u} + h\mathbf{v}) + f(\mathbf{x}_0 - h\mathbf{u} - h\mathbf{v})
\Big).
\end{aligned}
\]
By Taylor expansion, these approximations incur truncation error $O(h^2)$.

Employing Taylor expansions, we estimate the truncation error in these approximations:
$$
f(\mathbf{x}_0 \pm h \mathbf{u}) = f(\mathbf{x}_0) \pm h \nabla f \cdot \mathbf{u} + \frac{h^2}{2} \mathbf{u}^{\!\top} H_f \mathbf{u} + \frac{h^3}{6} D^3 f(\mathbf{u},\mathbf{u},\mathbf{u}) + O(h^4),
$$
$$
f_{uu} = \mathbf{u}^{\!\top} H_f \mathbf{u} + \underbrace{\frac{h^2}{12} \frac{\partial^4 f}{\partial \mathbf{u}^4}(\xi_1)}_{O(h^2)},
$$
$$
f_{vv} = \mathbf{v}^{\!\top} H_f \mathbf{v} + \underbrace{\frac{h^2}{12} \frac{\partial^4 f}{\partial \mathbf{v}^4}(\xi_1)}_{O(h^2)},
$$
$$
f_{uv} = \mathbf{u}^{\!\top} H_f \mathbf{v} + \underbrace{\frac{h^2}{12} \frac{\partial^4 f}{\partial u^2 \partial v^2}(\xi_2)}_{O(h^2)}.
$$
Theoretically, the finite-difference estimation of second derivatives and the second fundamental form has an approximation error of $O(h^2)$, demonstrating the accuracy and computational efficiency of our proposed method.

\section{Additional Quantitative Results}
\label{sec:validation}

\paragraph{Human-selected Examples}
On the cleaner 5\,MB subset, see Table \ref{tab:abc}, both finite difference variants exhibit strong performance across all metrics. Proxy-FD achieves the highest Normal Consistency (NC) and the lowest Chamfer Distance (CD), outperforming NeurCADRecon. Proxy-AD leads in F1 score, exceeding NeurCADRecon by over five points. These results highlight the ability of both proxy methods to generalize well on well-curated data, with Proxy-FD showing the most consistent geometric accuracy, and Proxy-AD providing the highest detection fidelity. Figure 2 in the main paper depicts a few selected models of the 5\,MB set. 

\paragraph{Pseudo-Random Examples}
In the more varied 1\,MB subset, see Table \ref{tab:abc}, where input meshes are less structured, Proxy-FD again leads in geometric accuracy with the lowest CD and a strong NC, closely trailing NeurCADRecon. F1 scores remain competitive, with Proxy-FD and Proxy-AD closely rivaling NeurCADRecon. These findings demonstrate that even with varied input point clouds, both Proxy-FD and Proxy-AD deliver reconstructions that are either on par with or superior to NCR, and significantly outperform DiGS and NSH across all metrics.

\paragraph{Efficiency}
In terms of computational efficiency, see Table \ref{tab:time_quant_eval}, NeuralSingularHessian remains the fastest method overall with a convergence time, owing to its minimal iteration cost. However, Proxy-FD achieves the fewest convergence iterations and maintains a competitive total runtime, outperforming all other methods except NeuralSingularHessian. Proxy-AD offers a strong trade-off with the second-lowest iteration count and runtime, while consuming modest GPU memory. Compared to NeurCADRecon's long runtime and high memory footprint, both Proxy methods are significantly more efficient. This demonstrates that Proxy-FD and Proxy-AD not only offer high-quality reconstructions but do so with favorable resource demands, making them well suited for scalable and time-sensitive applications.

Varying the weight of the FD-NCR weight $\lambda_{\text{FD}}$ shows stable reconstructions across a wide range (Table~\ref{tab:ablation_weights}, Fig, see paper). Even small weights improve developability and quality.

\begin{figure*}[t]
    \centering
    \includegraphics[width=0.99\linewidth,trim=50 0 50 0,clip]{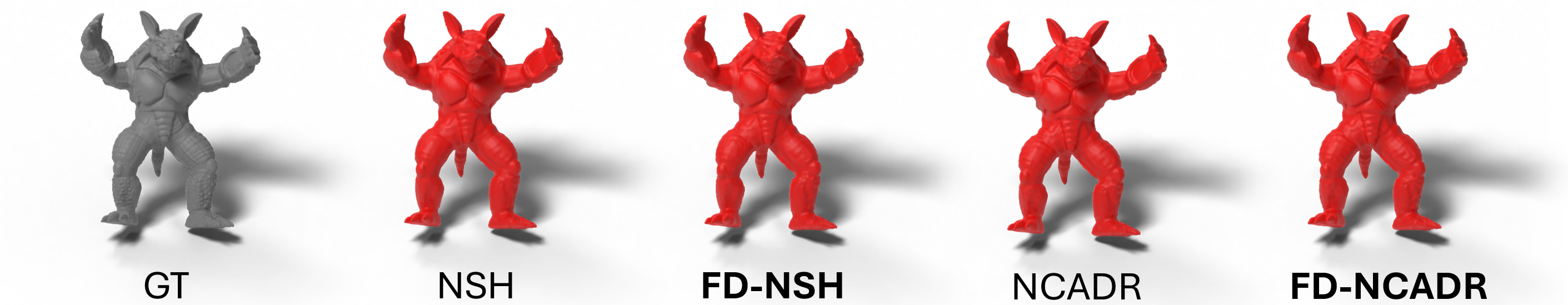}
    \caption{Reconstruction results on Armadillo. FD-NCR and NeurCADRecon give smooth, complete surfaces, NSH show more artifacts.}
    \label{fig:armadillo}
\end{figure*}

\paragraph{Generalization to Non-CAD}
On the non-CAD model (Armadillo), FD-NCR reconstructs smooth, topologically correct surfaces comparable to NRC but at about half the runtime (Fig.~\ref{fig:armadillo}, Table~\ref{tab:armadillo}). Fine detail is preserved to the same extend as the baselines.

\begin{table}[t]
  \centering
  \small
  \caption{Quantitative results on Armadillo reconstruction.}
  \label{tab:armadillo}
  \begin{tabularx}{\columnwidth}{l *{4}{>{\centering\arraybackslash}X}}
    \hline
    Method & NC~$\uparrow$ & CD~$\downarrow$ & F1~$\uparrow$ & Time [s] \\
    \hline
    NCR (FD)   & 0.9784759 & 0.0022561 & 0.9898897 & 1015.53 \\
    NCR (base) & 0.9789649 & 0.0022634 & 0.9898299 & 1891.00 \\
    \hline
  \end{tabularx}
\end{table}

\begin{table}[t]
  \centering
  \small
  \caption{Quantitative results with different regularization weights.}
  \label{tab:ablation_weights}
  \begin{tabularx}{\columnwidth}{l *{4}{>{\centering\arraybackslash}X}}
    \hline
    Weight & NC~$\uparrow$ & CD~$\downarrow$ & F1~$\uparrow$ & Time [s] \\
    \hline
    0   & 0.984766 & 0.0038015 & 0.889413 & 1683.27 \\
    0.2 & 0.979177 & 0.0040342 & 0.926001 & 1462.14 \\
    0.4 & 0.981741 & 0.0040857 & 0.929235 & 1494.68 \\
    0.6 & 0.993273 & 0.0024407 & 0.970599 & 1524.86 \\
    0.8 & 0.993057 & 0.0024853 & 0.968210 & 1540.46 \\
    1   & 0.992991 & 0.0024358 & 0.971480 & 1550.66 \\
    3   & 0.993265 & 0.0024326 & 0.972560 & 1603.98 \\
    5   & 0.993028 & 0.0024475 & 0.970151 & 1825.94 \\
    7   & 0.993131 & 0.0024744 & 0.968655 & 1820.79 \\
    9   & 0.959857 & 0.0080776 & 0.870467 & 1812.14 \\
    \hline
  \end{tabularx}
\end{table}

\begin{table*}
\centering
\small
\caption{Quantitative results on the ABC dataset~\cite{Koch_2019_CVPR}. Evaluation is conducted on two resolution subsets (1\,MB and 5\,MB) using three metrics: Normal Consistency (NC), Chamfer Distance (CD), and F1 score (F1). For each metric's mean value, the better result is \textbf{bold}.}
\label{tab:abc}
\renewcommand{\arraystretch}{1.2}
\begin{tabular*}{\textwidth}{@{\extracolsep{\fill}}l|cccccc|cccccc}
\hline
\multicolumn{1}{c|}{} & \multicolumn{6}{c|}{1\,MB set} & \multicolumn{6}{c}{5\,MB set} \\
\multicolumn{1}{c|}{} & \multicolumn{2}{c}{NC~$\uparrow$} & \multicolumn{2}{c}{CD~$\downarrow$} & \multicolumn{2}{c|}{F1~$\uparrow$} & \multicolumn{2}{c}{NC~$\uparrow$} & \multicolumn{2}{c}{CD~$\downarrow$} & \multicolumn{2}{c}{F1~$\uparrow$} \\
\multicolumn{1}{c|}{} & mean & std. & mean & std. & mean & std. & mean & std. & mean & std. & mean & std. \\
\hline

NSH~\cite{Wang2023NeuralSingularHessianIN}
    & \textbf{93.93} & \textbf{7.31} & \textbf{2.74} & 5.96 & \textbf{91.91} & \textbf{21.30}
    & \textbf{96.59} & 6.60 & \textbf{5.52} & \textbf{5.92} & \textbf{84.45} & \textbf{20.21} \\

NSH-FD
    & 94.96 & \textbf{5.66} & 2.93 & \textbf{4.78} & 89.90 & 23.60
    & 95.20 & 6.63 & 8.78 & 9.75 & 74.27 & 30.62 \\

\hline

NCR~\cite{Dong2024NeurCADReconNR}
    & \textbf{93.71} & 7.40 & \textbf{2.65} & \textbf{3.76} & \textbf{90.63} & \textbf{20.23}
    & \textbf{96.17} & \textbf{5.53} & \textbf{5.90} & \textbf{4.86} & \textbf{79.66} & \textbf{22.86} \\

NCR-FD
    & 93.41 & 8.16 & 4.10 & 6.31 & 84.78 & 26.99
    & 94.80 & \textbf{6.58} & 8.49 & 9.72 & 72.91 & 31.77 \\

\hline
\end{tabular*}
\end{table*}

\begin{table*} %
\centering
\small
\caption{Comparison of iteration time (ms), convergence time (s)—computed as mean iteration time $\times$ number of iterations, scaled to seconds—and GPU memory usage. Within each column, the better is \textbf{bold}.}
\label{tab:time_quant_eval}
\begin{tabular*}{\textwidth}{@{\extracolsep{\fill}}l | cc c c c@{}}
\hline
& \multicolumn{2}{c}{Iter.\ time (s)} & Conv.\ iter & Conv.\ time (s) & GPU Mem.\ (GB) \\
& mean & std. & mean & mean & mean \\
\hline
NSH~\cite{Wang2023NeuralSingularHessianIN} & 6.11 & \textbf{0.06} & 9159 & 559.61 & 6.09 \\
NSH-FD & \textbf{4.04} & \textbf{0.06} & \textbf{8984} & \textbf{362.95} & \textbf{4.30} \\
\hline
NCR~\cite{Dong2024NeurCADReconNR} & \textbf{4.11} & \textbf{0.02} & 9519 & 391.23 & 6.06 \\
NCR-FD & 6.03 & 0.27 & \textbf{5494} & \textbf{331.29} & \textbf{4.03} \\
\hline
\end{tabular*}
\end{table*}

\bibliographystyle{ACM-Reference-Format}
\bibliography{related_work_bib,related_inrs}

\end{document}